\documentclass[runningheads]{llncs}
\usepackage[T1]{fontenc}
\usepackage{subcaption}
\usepackage{enumitem}

\usepackage{hyperref}
\usepackage{cleveref} %
\usepackage{booktabs}

\newif\ifdraft
\drafttrue

\ifdraft
\newcommand{\authorcomment}[3]{\textcolor{#2}{\noindent[#1: #3]}}
\newcommand{\todo}[1]{\textcolor{red}{TODO: #1\xspace}}

\newcommand{\todocite}[1]{\textcolor{OliveGreen}{TODOCITE: #1}}
\newcommand{\owner}[1]{\textcolor{red}{\noindent[Owner: \textbf{#1}]}}
\newcommand{\budget}[1]{\textcolor{orange}{\noindent\textbf{Budget: #1}}}
\else
\newcommand{\authorcomment}[3]{}
\newcommand{\todo}[1]{}

\newcommand{\todocite}[1]{}
\newcommand{\owner}[1]{}
\newcommand{\budget}[1]{}
\fi

\newcommand{\ignore}[1]{}

\newcommand{\parx}[1]{\noindent\textbf{#1}}

\usepackage{xspace}
\usepackage[dvipsnames]{xcolor}
\usepackage{graphicx}

\newcommand*{\eg}{e.g.,\@\xspace}
\newcommand*{\ie}{i.e.,\@\xspace}
\newcommand*{\et}{et al.\@\xspace}
\newcommand{\zmap}{ZMap\xspace}
\newcommand{\zmapsix}{ZMapv6\xspace}
\newcommand{\zgrab}{ZGrab2\xspace}

\newcommand{\PP}[1]{
	\vspace{2px}
	\noindent{\bf \IfEndWith{#1}{.}{#1}{#1.}}
}

\usepackage{multirow}
\usepackage[normalem]{ulem}
\useunder{\uline}{\ul}{}

\usepackage[breakable, skins]{tcolorbox}

\newcommand{\takeaway}[1]{
\begin{tcolorbox}[title=\textit{Takeaway},enhanced,breakable]
    #1
\end{tcolorbox}
}
\renewcommand{\takeaway}[1]{\emph{\textbf{To summarize: }{#1}}}
 
\begin{document}
\title{A First Look At NAT64 Deployment In-The-Wild}
\author{Amanda Hsu\inst{1}
    \and
    Frank Li\inst{1}
    \and
    Paul Pearce\inst{1}
    \and
    Oliver Gasser\inst{2}
}

\institute{Georgia Institute of Technology
    \and
    Max Planck Institute for Informatics
}

\maketitle              %
\begin{abstract}
\begin{enumerate}
    IPv6 is a fundamentally different Internet Protocol than IPv4, and
    IPv6-only networks cannot, by default, communicate with the IPv4 Internet.
    This lack of interoperability necessitates complex mechanisms for
    incremental deployment and bridging networks so that non-dual-stack systems
    can interact with the \emph{whole} Internet.  NAT64 is one such bridging
    mechanism by which a network allows IPv6-only clients to connect
    to the entire Internet, leveraging DNS to identify IPv4-only networks,
    inject IPv6 response addresses pointing to an internal gateway, and
    seamlessly translate connections.  To date, our
    understanding of NAT64 deployments is limited; what little information
    exists is largely qualitative, taken from mailing lists and informal
    discussions.

    In this work, we present a first look at the active measurement of
    NAT64 deployment on the Internet focused on deployment prevalence, configuration, and security. We seek to measure NAT64 via two distinct
    large-scale measurements: 1) open resolvers on the Internet, and 2) client
    measurements from RIPE Atlas.  For both datasets, we broadly find that
    despite substantial anecdotal reports of NAT64 deployment,
    \emph{measurable}  deployments are exceedingly sparse.  While our measurements
    do not preclude the large-scale deployment of NAT64, they do point to
        substantial challenges in measuring deployments with our existing
        best-known methods.
Finally, we also identify problems in NAT64 deployments, with gateways not following the RFC specification and also posing potential security risks.
\end{enumerate}

\end{abstract}
\section{Introduction}

The modern Internet is a mix of both IPv4 and IPv6 hosts.  With the exhaustion
of the IPv4 address space, network providers are increasingly turning to IPv6 for
expansion and new deployments.  This increase in IPv6 adoption is measurable
and accelerating; as of November 2023, Google estimates that over 40\% of its users connect over native
IPv6, up from less than 1\% a decade ago~\cite{google_ipv6_stat}. 
However, this migration has been fraught, complex, and slow, and there is no built-in interoperability between IPv4 and IPv6 protocols.

To bridge IPv6-centric networks into the IPv4-only Internet, transition
protocols have been proposed~\cite{transition_mechs_2005_rfc4213}, deployed~\cite{464xlat_tmobile}, and
deprecated~\cite{6to4_anycast_deprecated_rfc7526}. 
The recent acceleration of IPv6 adoption has brought renewed interest in
understanding how networks can ease this transition. Among currently
deployed tools is NAT64, a mechanism that allows IPv6-only systems to access
IPv4-only Internet resources. NAT64 leverages DNS64 resolvers to map IPv4
addresses into IPv6 addresses; when an IPv6-only client issues a DNS
request for a domain that only contains an A record, a resolver
(implementing the companion DNS64 protocol) will insert synthetic AAAA records
mapping to a ``gateway'' IPv6 address that will proxy the connections
from IPv6 to IPv4, in a mechanism akin to traditional Network Address
Translation (NAT).  This mechanism is economically efficient; it allows
operators to keep the edges of their networks, new deployments, and clients, to
be IPv6 only; only gateways need to have support for IPv4, without sacrificing
customer experience, or needing IPv4 NAT or IPv4
management~\cite{464xlat_whitepaper}.
Moreover, this technique also enables measurement
opportunities, as the synthetic AAAA records can be identified to locate and
understand NAT64 deployments.

Anecdotally, substantial NAT64 deployments are reported. The 464XLAT
protocol~\cite{464xlat_rfc6877} integrates NAT64 and is deployed by T-Mobile
in the United States~\cite{464xlat_tmobile}, and is reported to be deployed in
multiple other networks including Sprint, Telstra, and Deutsche
Telekom~\cite{ietf_mailing_list_v6ops}.  Additionally, China Mobile has reported its
intention to use it to maintain connectivity to both IPs as it deploys
IPv6~\cite{china_rfc6219}.  Alongside these deployments, concerns for
misconfiguration in transition mechanisms that could impact clients' Internet
accessibility arose~\cite{nat64_checker}.  

To date, NAT64 has not been empirically explored across the Internet.  In this work we provide a first look at NAT64 deployment
in-the-wild, seeking to understand the prevalence of deployment, configuration
issues, and security shortcomings. We achieve this by issuing DNS queries designed to
elicit NAT64 responses across both real-world and control IPv4-only domains
and explore the responses, and the properties of the machines at the returned
addresses.

Our contributions include:
\begin{itemize}[nosep, leftmargin=*]
    \item \textbf{NAT64 deployment analysis:}
        We present the first large-scale NAT64 analysis, finding 2,021 deployments across 262 Autonomous Systems (ASes).
    \item \textbf{Configuration:}
        We find 60.7-100\% of NAT64 configurations are correctly configured when embedding IPv4 addresses within synthetic AAAA records.
    \item \textbf{Security:}
    	Finally, we identify potential security issues, with 1.1--26.6\% of NAT64 gateways being publicly accessible.
\end{itemize}

\section{Background and Related Work}
\label{sec:background}

In this section, we present an overview of transition mechanisms in general and NAT64 specifically and elaborate on existing work in NAT64 measurements.

Transition mechanisms can be categorized in three ways: single transition, double transition, and tunneling. 
A single transition mechanism translates IPv4/6 traffic to IPv6/4 traffic, and vice versa. 
A double transition mechanism translates IPv4/6 to IPv6/4 and then back to IPv4/6. 
A tunnel encapsulates IPv4/6 in IPv6/4 \cite{6to4_rfc3056}.
NAT64 is an example of a single transition whereby IPv6 traffic is translated to IPv4 on outbound connections, and then back again for responses.

\label{background:nat64}

\parx{NAT64 Background:}
NAT64 is a transition mechanism that allows an IPv6-only host to access
IPv4-only Internet resources~\cite{rfc6146_nat64}.  This mechanism consists of
two parts: A DNS64-enabled DNS resolver and a NAT64 gateway.
For each domain requested by an IPv6 client, the DNS64
resolver returns a AAAA record.  If the requested domain does not have a AAAA
record and only an A record, the DNS64 resolver will create a synthetic AAAA
record pointing to and designed to be read by a NAT64 gateway.
This synthetic AAAA record may use a reserved NAT64 prefix (\texttt{64:ff9b::/96} or \texttt{64:ff9b:1::/48}~\cite{newer_rfc8215}), public IPv6 space, or other private address space.
The AAAA record embeds the IPv4 address from the A record in the last 32 bits of the
IPv6 address, which enables the gateway to translate traffic to and from the IPv4 resource to the IPv6 client.
While largely used privately, a number of public NAT64 and DNS64 services are publicly
available~\cite{nat64.net}.
 Although DNS64 is specifically designed to be used with NAT64, both
of these systems do not need to be deployed in the same network (\eg Google offers a free DNS64 service, but not NAT64~\cite{google_public_dns64}).

\parx{NAT64 Deployments:}
The stateless nature of NAT64 makes it apt and scalable in many
networks~\cite{rfc6144}.  Additionally, it may be built into other transition mechanisms such as 464XLAT~\cite{464xlat_rfc6877},
which allows for IPv4 and IPv6 networks alike to share a part of their
infrastructure, again, adding to the network scalability.  T-Mobile is known
for deploying this mechanism in the United States~\cite{464xlat_tmobile}.
In addition to scalability, both of these mechanisms are also more cost-efficient in large networks~\cite{464xlat_whitepaper}.
Moreover, NAT64 and DNS64 have been reported to be deployed by a number of other large network providers such as Deutsche Telekom in Germany~\cite{ietf_mailing_list_v6ops}. 
NAT64 deployment may have also been motivated by Apple's requirement for mobile iOS applications to have IPv6 connectivity to use the app store~\cite{apple_ipv6_req}. 
App developers report leveraging local NAT64 deployments in their testing~\cite{nat64_tayga_testing}.
Finally, we highlight the need for NAT64 to be able to access popular domains.  Just over
30\% of domains in the Cisco Umbrella top list are available via IPv6~\cite{employees_aaaa_stats}.
Therefore, to maintain accessibility to a large fraction of the
Internet, IPv6 networks must either deploy some form of IPv4 connectivity or
use a transition mechanism such as NAT64. 

\parx{Prior NAT64 Measurements:}
Prior work on transition mechanisms has focused on
performance~\cite{transition_performance_2021} and
classification~\cite{transition_classification_2019}.  Kristoff~\et measured
deprecated tunneling transition mechanisms and found many to still exist in ISP
backbones~\cite{plight_at_the_end_kristoff_21}.
Moreover, Zorz~\et deployed point-wise public testing infrastructure focused on website compatibility~\cite{nat64_checker}, which is unfortunately no longer operational. 
To the best of our knowledge, we present the first measurement study of NAT64 on the public
Internet. 

\section{Methodology}

\begin{figure}[t]
	\centering
	\includegraphics[width=.9\columnwidth]{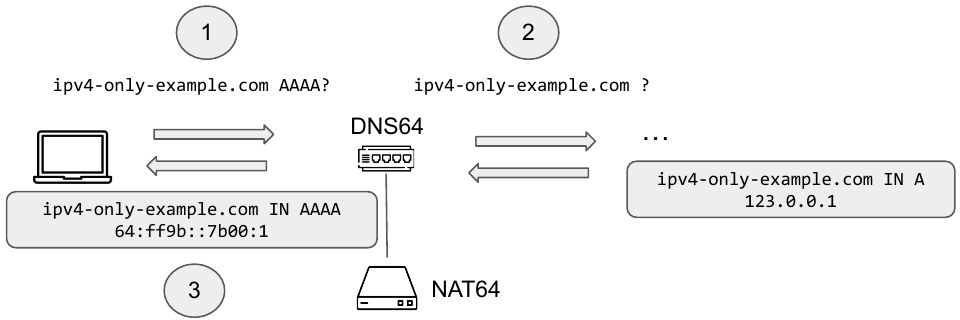}
    \caption{An overview of our methodology. We request a AAAA record for an IPv4-only domain (1), the DNS64 resolver resolves the domain and only finds an A record (2), and returns a synthetic AAAA record (3).}
	\label{fig:method}
\end{figure}

Our methodology (see \Cref{fig:method}) is centered around requesting a AAAA record for a domain where we know that the authoritative nameserver only answers with an A record.
In step (1) we send AAAA queries for IPv4-only domains to different public resolvers.
In step (2) the DNS64 resolver will then try to resolve A and AAAA for the requested domain.
In step (3) if only an A record is returned, the DNS64 resolver creates a synthetic AAAA record to be returned to the client.
The returned address is usually in the reserved NAT64 prefix (\texttt{64:ff9b::/96}). The IPv4 address is embedded in the last 32 bits as required according to the NAT64 RFC~\cite{rfc6146_nat64}. 
We expand on each part of this method.

\parx{Choosing Target Domains:}
We sample four domains from the Tranco top 1M list~\cite{tranco_LePochat2019} \footnote{We use the Tranco list~\cite{tranco_LePochat2019} generated on October 5, 2023, available at \url{https://tranco-list.eu/list/W9P79}.}
and measure them for AAAA and A records using ZDNS~\cite{zdns}.
We ensure that each domain is online by measuring it for HTTP and HTTPS with \zgrab~\cite{zgrab2} so that we can later perform application-layer measurements of the NAT64 gateways.
To avoid censorship, we filter out domains in the Citizenlab global censorship list~\cite{citizenlab_testlist}, manually discard domains related to social media or journalism, and ensure that none elicit censorship behavior from China's Great Firewall (GFW) as observed in prior work~\cite{zirngibl2022rusty,steger2023targetacquired}.
To avoid overloading authoritative nameservers, we randomly sample from the first 1k domains in Tranco, as these likely expect higher traffic volumes. 
The domains are:
\texttt{azure.com}, 
\texttt{cloudflare.net},
\texttt{ntp.org},
and \texttt{webex.com}
We also measure \texttt{github.com}, as it is a popular domain that only supports IPv4 and is typically not censored.

\parx{Identifying Public Resolvers:}
To identify public IPv6 resolvers, we use the IPv6 hitlist~\cite{clusters_in_the_expanse} snapshot from October 18, 2023, containing 279.1k public IPv6 resolvers.
To identify public IPv4 resolvers, we use the Censys dataset~\cite{censys} snapshot from October 16, 2023, containing 1.9M public IPv4 resolvers.

\parx{Resolver Measurements:}
After identifying target domains and public resolvers, we use \zmap~\cite{zmap} and \zmapsix~\cite{clusters_in_the_expanse} to request AAAA records for each of our domains. 
We also request an A record for each domain-resolver pair responding with a valid AAAA record to validate that it correctly embeds the IPv4 address according to the NAT64 RFC~\cite{rfc6146_nat64}, as explained in Section~\ref{sec:background}.
To avoid misclassifications that may have arisen due to caching for domains with multiple valid A records, we also check the IPv6 address for embeddings of other valid A records for the domain.

\parx{RIPE Atlas Measurements:}
To understand client DNS64 deployment, we use RIPE Atlas probes~\cite{ripe_atlas_platform} %
to measure AAAA records for our chosen domains using their local resolver.
We run measurements from all 5,574 probes that are online and have IPv6 connectivity\footnote{We filter for the probe tag \texttt{system-ipv6-works}, which indicates that the RIPE Atlas system confirmed the probe's IPv6 connectivity.}.
In addition, we also run IPv4 measurements on 5,472 probes which are also IPv4-connected\footnote{For probes also tagged with \texttt{system-ipv4-works}, we run the measurement over IPv4.}.
As with our resolver measurements, we issue AAAA as well as A queries to check for valid embeddings of IP addresses.

\parx{Application-Layer Measurements Through NAT64 Gateways:}
With the public IPv6 addresses in DNS answers from the measurements towards our sampled domains,
 we make HTTPS requests with \zgrab~\cite{zgrab2}. 
 We make these measurements from a university network in Germany. 
 We make these measurements by using SNI to specify the domain name alongside the IPv6 address. 
From this, we determine whether the gateway operating alongside the DNS64 resolver is publicly accessible. 
We perform these measurements with all public IPv6 addresses
 in case the gateway is operating as a NAT64, but keeps some state or uses SNI that is not aligned with the RFC~\cite{rfc6146_nat64}. 

To determine whether we can access the website content through the gateway, we compare our results to control measurements by measuring the results of HTTP and HTTPS requests to the addresses on the domain's A records. 
For our HTTPS measurements, we classify a certificate as ``correct'' if it has the requested domain on it. 
For our HTTP measurements, we classify a response as ``correct'' if it matches the response body or the redirect from a control measurement.

\parx{Identifying NAT64 Gateways:}
To identify the ``IPv4-side'' of NAT64 gateways,
we host an authoritative nameserver for an IPv4-only domain.
Then, we send AAAA queries for our domain from RIPE Atlas probes to their local resolvers.
We then run TCP/80 pings to our domain from the probes to identify the probe's NAT64 gateway by recording the IPv4 address reaching our webserver. 
Due to platform restrictions, we cannot measure probes that return private addresses outside of the reserved NAT64 prefix. The platform restricts measurements towards arbitrary private IP space besides select special-use prefixes. 

\parx{Ethical Considerations:}
\label{sec:ethics}
We follow ethical scanning guidelines in our measurements \cite{zmap,dittrich2012menlo,partridge2016ethical}.
Specifically, we use a blocklist, maintain public web pages describing our scanning activity, and scan at low rates (less than 100 pps). 
We additionally ensure we do not request any potentially censored domains. 
We emphasize that, at most, we are sending 10 DNS requests to each resolver.
Finally, we sample our domains from popular Tranco domains (the least popular domain is on rank 132), as these generally expect a higher volume of traffic at the authoritative.
During our measurements, we did not receive any complaints.

\parx{Limitations:}
Our study is limited to publicly available resolvers and RIPE Atlas probes. 
Due to the nature of the RIPE Atlas measurement platform, our client measurements may not represent NAT64 deployment at large. 
As described in Section~\ref{background:nat64}, NAT64 seems to be predominantly deployed in mobile networks, which are less likely to be in RIPE Atlas, although some are~\cite{ripe_atlas_docs}.
We also measure for popular domains, but it is possible that resolvers respond differently to less-popular domains.
Despite these limitations, our methodology is suitable to provide a lower bound on NAT64 deployment. 

\section{Datasets}
\label{sec:dataset}

In this section, we describe our data and how we find NAT64 deployments.

\parx{Resolver Responses:}
Out of 1.9M
IPv4 resolvers measured, 1.0--1.1M respond to our scans, depending on the domain requested (52.6--57.9\%).
Out of 279.1k  
IPv6 resolvers measured, 265.0k-267.2k (94.9--95.7\%)
respond, also depending on the requested domain. 
We attribute the lower response rate of IPv4 to a higher churn rate compared to IPv6.
Of the responsive IPs, 0.68--0.82\% of answering IPv4 resolvers and 0.13--0.15\% of answering IPv6 resolvers respond with answers to our AAAA queries, hinting at possible DNS64 deployments.
Moreover, 0.016--0.029\% of IPv4 resolvers send malformed responses (\eg uninterpretable hexadecimal characters).
Over half of all IPv4 resolvers (52.9--55.5\%) respond with DNS responses with no answer (\ie an \texttt{ancount} of 0), showing that these are not DNS64 resolvers, however less than 10\% (9.4--9.5\%) of IPv6 resolvers respond this way.
We find that 6,335 IPv4 resolvers and 309 IPv6 resolvers consistently respond with DNS answers across all domains, the majority of which are of type AAAA (70.3\% IPv4, 88.0\% IPv6).

Despite requesting a AAAA record, 20.9--24.3\% of answering IPv4 resolvers and
1.2--1.3\% of answering IPv6 resolvers return an A record, signaling a resolver
misconfiguration or perhaps that we 
reached DNS forwarders in residential gateways that respond with cached values, not actual resolvers~\cite{client_side_dns_schomp2013}.
For the remainder of our analysis, we focus on the answers we receive in AAAA responses. 
Although we see a relatively uniform distribution of the types of responses across resolvers, certain resolvers elicit different quantities of responses from resolvers in certain ASes, especially in requests for \texttt{webex.com} in resolvers in China. 
We expand on the responses of resolvers in more detail in \Cref{app:resolver_responses}.

\parx{Filtering AAAA Answers:}
\label{sec:filtering_aaaa}
Next, we identify which answers from resolvers are indicators for DNS64 deployments by filtering AAAA answers.
Specifically, we filter out resolvers if they (1) answer with the same IP independent of the requested domain, (2) answer with an invalid IPv6 address (\eg that starts with \texttt{::}), or (3) answer with anything other than an IPv6 address in the \texttt{answer} field of the DNS response (domains, IPv4 addresses, and miscellaneous values). We characterize the filtered addresses in Appendix~\ref{app:resolver_answers}.

We filter out these 172 IPv6 resolvers and 2,454 IPv4 resolvers from future analysis.
We continue our analysis with the remaining 240 IPv6 and 6,201 IPv4 resolvers that respond consistently to \textit{any} domain query after our filtering.

\parx{RIPE Atlas Probes:}
We find 39 probes that respond across all domains, one of which answers with an NS record that we filter out from our results. 
28 have IPv4 and IPv6 connectivity, and 11 are IPv6-only probes. 
\footnote{We find that in total 102 RIPE Atlas probes are IPv6-only. We leave investigating if and how these probes can reach IPv4-only resources to future work.}
We find that nine probes answer inconsistently across domains.

We note that five of the 39 probes that respond across all domains are from AS 3320, Deutsche Telekom, a Tier-1 ISP from Germany, and one is from AS 2856, British Telecommunications. 
Both of these ASes have been reported as using 464XLAT and NAT64~\cite{ietf_mailing_list_v6ops}.
However, we note that ASN 3320 and AS 2856 host a total of 856 and 231 probes, respectively. 
Therefore, the probes we identify as using NAT64 are a small part of these networks. 
Overall, these 39 probes cover 32 unique IPv6 ASes and 22 unique IPv4 ASes.

\parx{Measurements Towards Our Nameserver:}
\label{sec:inside_measurements}
Based on our RIPE Atlas measurements for the five popular domains, we additionally conduct measurements towards our webserver from 47 RIPE Atlas probes that answered with valid NAT64 responses to any domain.

We receive responses from 33 of these probes when requesting AAAA records for our domain. 
Four respond with public IPv6 addresses, one responds with a private address, and the rest respond with addresses in the special use NAT64 prefix.
Due to limitations in RIPE Atlas, we are unable to measure from a probe to private addresses.
We do, however, conduct measurements from the remaining probes towards the IPv6 addresses in their respective DNS responses.

As a result, we identify NAT64 gateway addresses on 25 probes.
One probe did not participate in our measurement, and the remaining six probes did not complete our measurement successfully.
Over all of these probes, we identify 21 unique IPv4 gateway IP addresses in 18 ASes.
\section{NAT64 Deployment}
\label{sec:resolver_results}

In this section, we present the results of our resolver measurements and RIPE Atlas measurements towards our sampled popular domains as well as towards our own nameserver.
Specifically, we answer the following questions in this section:

\noindent\textbf{Deployment and demographics:}
	\begin{itemize}[nosep]
		\item What is the prevalence of NAT64 in the wild?
		\item Where are NAT64 gateways deployed?
		\item Where are DNS64 resolvers deployed?
	\end{itemize}
\textbf{Configuration:}
	\begin{itemize}[nosep]
		\item How are NAT64 gateways configured?
		\item Do NAT64 gateways correctly embed IPv4 addresses in NAT64 addresses?
	\end{itemize}
\textbf{Security:}
	\begin{itemize}[nosep]
    	\item Are NAT64 gateways publicly accessible?
	\end{itemize}

\subsection{Deployment and Demographics}

\parx{Prevalence:}
Out of 1.9M public IPv4 resolvers measured, 1,873 deploy NAT64 across all of our domain measurements (0.1\%). 
For IPv6, we identify 110 out of the 279.1k IPv6 resolvers deploying NAT64 across all of our domain measurements (0.04\%).
Out of 5,574 RIPE Atlas probes measured, 39 deploy NAT64 and consistently respond to our measurements (0.7\%).

\takeaway{NAT64 is not deployed widely across the Internet in public resolvers and the networks of RIPE Atlas probes.}

\begin{figure}[tb]
	\includegraphics[width=\columnwidth]{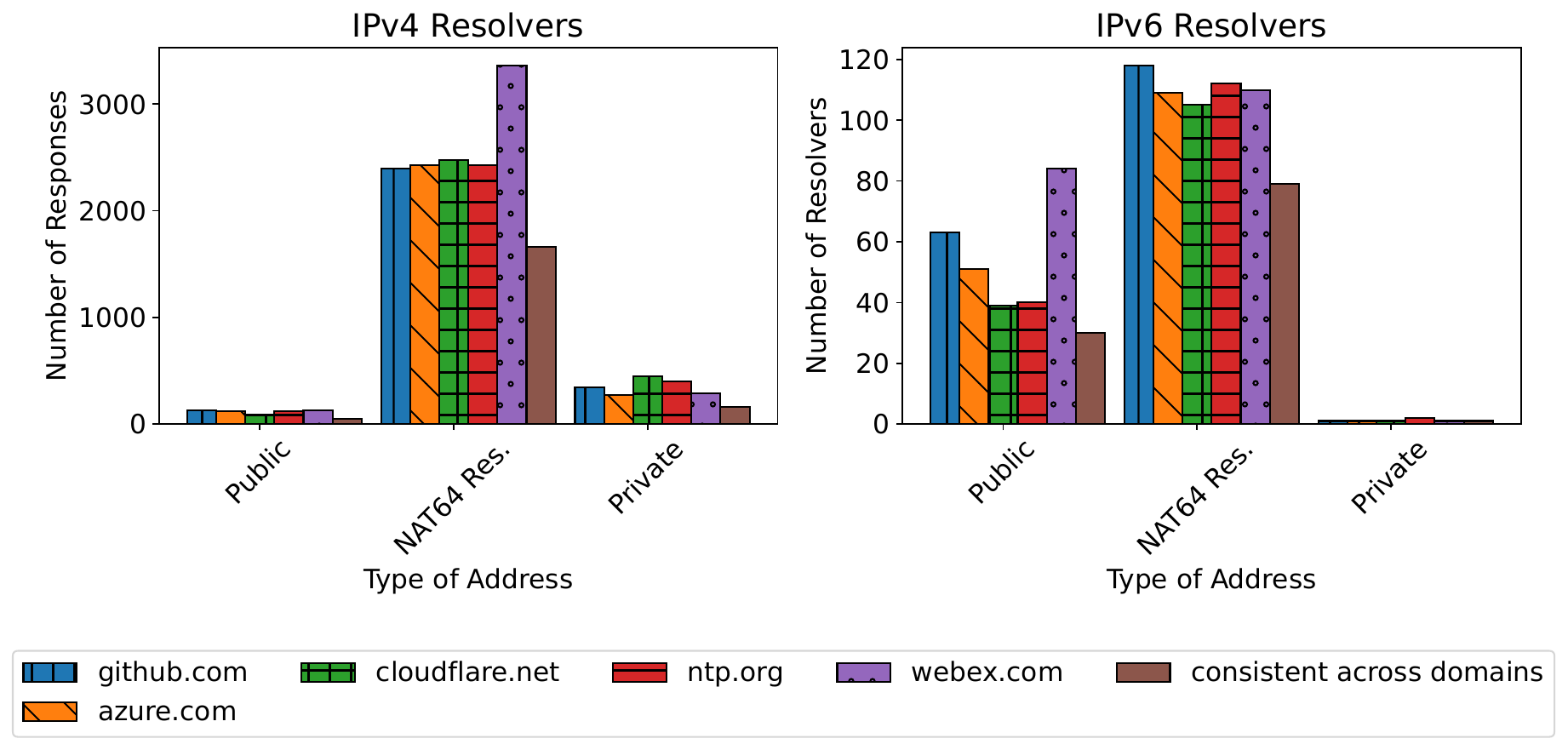}
	\caption{Resolver responses after filtering out invalid responses. We highlight that the inconsistencies around measurements for webex.com remain.}
	\label{fig:resolver_legit_answers}
\end{figure}

\parx{Gateway Location:}
Next, we investigate where NAT64 gateways are deployed.
We present an overview of the types of addresses returned by DNS64 resolvers in Figure~\ref{fig:resolver_legit_answers}. 
We highlight that the majority of DNS64 resolvers use NAT64 gateways that use the assigned NAT64 special use prefix. 
Additionally, in each case, a small number of resolvers use other private IP address space.
However, we also find cases in which resolvers return public IPv6 addresses.
In these cases, we compare the AS of the returned IPv6 address to that of the resolver. 
In IPv4, we find that 18 out of the 44 IPv4 resolvers responding with public IPv6 addresses are located in the same respective AS.
In IPv6, this is the case for 21 out of the 30 IPv6 resolvers responding with public gateway addresses. 
We manually confirm that in all but two of these cases across both IPs, the ASes are not siblings, but the gateway AS is a network services provider. 

Of the 38 RIPE Atlas probes that respond consistently, 28 of these probes respond with IPv6 addresses in the NAT64 special use prefix, and nine respond with IPv6 addresses that are publicly routable.
One probe responds with a private address in the unicast prefix (\texttt{fc00::/7}~\cite{iana_special_use}).
These responses are consistent across all domains for all but six probes.
When probing our own authoritative nameserver, out of the 33 probes that respond to our AAAA queries, four respond with public IPv6 addresses, one responds with a private address, and the rest respond with addresses in the NAT64 special use prefix.
Out of the 25 probes that complete the measurement towards the gateway,
we find that seven gateway IPs are the same as the probe's assigned public IPv4 address. 
This strongly indicates that the probe is behind a NAT that also acts as a NAT64 gateway. 
Overall, we find that 20 out of 25 gateway IPs are in the same AS as the probe. 
Again, we confirm that the disagreeing ASes are not siblings, but the gateway's AS is always from a network services provider.

\takeaway{The NAT64 deployment strategy varies, but most are only accessible through private networks that utilize the special use NAT64 prefix.
The gateways are likely to exist in the same NAT that the client's IPv4 address is behind, in the same network, or in a different network entirely.
}

\begin{table}[tb]
	\centering
    \resizebox{\textwidth}{!}{%
	\begin{tabular}{cccccccc}
		\toprule
		& \textbf{\#} & \textbf{github} & \textbf{azure} & \textbf{cloudflare} & \textbf{ntp} & \textbf{webex} \\
		\midrule
		\textbf{IPv4} & 1 & Meditelecom & Tata Tel. & Meditelecom & Meditelecom  & China Edu. \\
		& 2 & Tata Tel. & Meditelecom & China Edu. & Tata Tel. & Meditelecom  \\
		& 3 & Tata Tel. & Tata Tel. & Tata Tel. & China Edu. & Tata Tel. \\
		& 4 & China Telecom & China Edu. & Tata Tel. & Tata Tel. & Tata Tel. \\
		& 5 & Yettel Hun. & Yettel Hun. & China Tel. & China Tel. & China Tel. \\
		\midrule
		\textbf{IPv6} & 1 & China Tel. & China Tel. & China Tel. & China Tel.  & China Tel. \\
		& 2 & Hurricane El. & Hurricane El. & Hurricane El. & Hurricane El.  & Akamai \\
		& 3 & Akamai & Cernet & China Unicom & China Unicom & Hurricane El. \\
		& 4 & Cernet & Giginet & Cernet & Cernet & China Unicom \\
		& 5 & Giginet & Mythic Beasts & Mythic Beasts & Giginet & Cernet \\
		\bottomrule
	\end{tabular}
    }
	\caption{The top 5 ASes of DNS64 resolvers across IPv4 and IPv6 for different domain measurements. 
	Tata Teleservices has two different ASes that both appear in the top 5 ASes in all IPv4 resolver measurements.}
	\label{table:asn_resolvers}
\end{table}

\parx{Resolver Location:}
To understand in which networks DNS64 resolvers are deployed, we analyze the ASes of the resolver IPs. 
We show the top 5 ASes of NAT64 resolvers that respond to each domain in Table~\ref{table:asn_resolvers}. 
Deployment is relatively concentrated; 49.0--58.3\% of IPv4 and 51.1--56.4\% of IPv6 resolvers are in the top 5 ASes, despite covering 336--358 and 49--60 ASes, respectively.

Across all IPv4 resolvers, the most common AS of the resolvers is AS 4538 (China Education Network), followed by AS 134540 (Tata Teleservices). 
Curiously, the top AS that IPv4 DNS64 resolvers are concentrated in varies depending on the domain requested. 
We attribute this to our hypothesis about caching and prioritization of different domains, as discussed above.
Across IPv6 resolvers in all cases except for one domain, the most common AS is AS 4134 (China Telecom), followed by AS 6939 (Hurricane Electric). 
We can therefore conclude that legitimate NAT64 deployment is most common in China. 
We manually investigate the \textit{type} of business of each AS and
find that out of the 13 unique ASes across all IPv4 and IPv6 resolvers deploying DNS64, four are mobile service providers, four are network service providers, two are Chinese university networks, two are hosting providers, and one is a CDN.
For the 39 RIPE Atlas probes, which we identified as using NAT64, five probes are located in AS3320 (Deutsche Telekom, a German ISP) and four are in AS 2027 (MilkyWan, a French ISP).

\takeaway{DNS64 resolvers are concentrated in mobile networks, network service providers, and Chinese networks.}

\subsection{NAT64 Gateway Configuration}

\begin{figure}[tb]
	\includegraphics[width=\columnwidth]{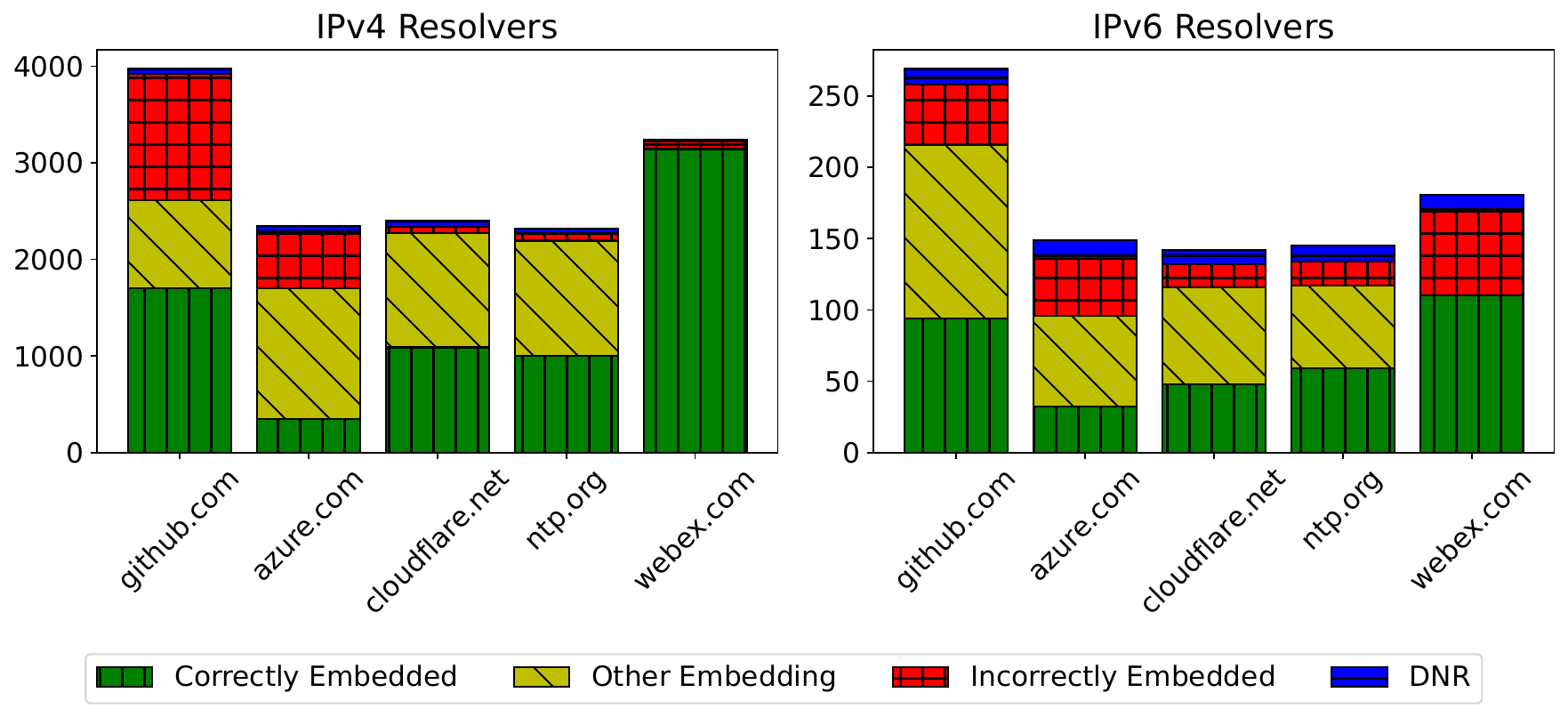}
	\caption{
		An overview of the answers for domains that correctly and incorrectly embed the corresponding A record. 
		We note that github.com also has more answers in general than other domains. ``Other Embedding'' refers to resolvers that respond with a correct embedding for the requested domain, but not the same IPv4 address that the resolver responds with when the domain's A record is requested.
		``DNR'' refers to the resolvers that did not respond to our measurements for A records. 
	}
	\label{fig:resolver_a_embeddings}
\end{figure}

Next, we analyze the embedding of IPv4 addresses within the returned AAAA records.
Figure \ref{fig:resolver_a_embeddings} shows the number of resolvers that correctly and incorrectly embed the respective A record into the AAAA record. 
When we check other IPv4 addresses from other valid A records for the respective domain, we find that the embeddings are generally correct.
For IPv6 resolvers, we find that for all domains except for \texttt{webex.com}, 61.3--80.9\% of the incorrectly embedded AAAA answers are in this category. 
For \texttt{webex.com}, we find that none of the incorrectly embedded answers fall into this category (\ie if an answer is incorrectly embedded, it never embeds a different IPv4 address for the domain). 
For IPv4 resolvers, we find that across all domains except for \texttt{webex.com}, 41.2--95.1\% of incorrectly embedded answers fall into this category. 
For \texttt{webex.com}, we once again find a much lower rate of this occurring at 2.4\%.

We conclude that the majority of our identified resolvers have legitimate NAT64 deployments that follow the specifications in the RFC~\cite{rfc6146_nat64}. 
Discrepancies in the embeddings are likely due to the caching of different A records at different times. 
We hypothesize that the differences in measurements for \texttt{webex.com} are likely due to it being cached much closer to the resolvers we are measuring.

\begin{figure}[tb]
	\centering
	\includegraphics[width=\columnwidth]{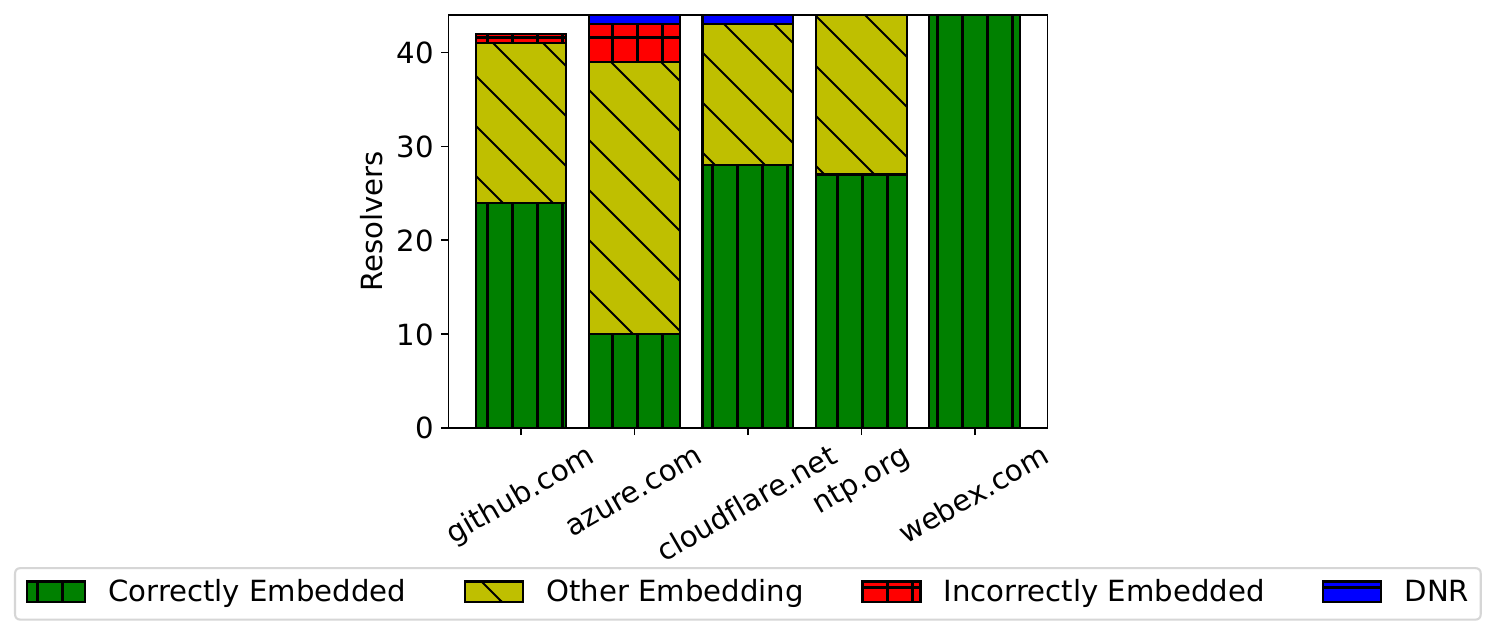}
	\caption{The results of analyzing the embeddings of the IPv4 addresses in the AAAA records returned by RIPE Atlas probes across domain measurements.}
	\label{fig:ripe_atlas_embeddings}
\end{figure}

For our RIPE Atlas measurements, we compare the IPv6 addresses on AAAA records to the IPv4 addresses on the A records from the same resolver; see \Cref{fig:ripe_atlas_embeddings}. 
 Although we identify in all domains (excluding \texttt{webex.com}) that there is a high rate of incorrectly embedded IPv4 addresses, we find that 91.7--100\% of these IPv6 addresses embed a valid IPv4 address for an A record of the domain. 

\takeaway{Overall, DNS64 resolvers correctly embed IPv4 addresses in synthetic AAAA records, but up to 39.2\% of resolvers use an incorrect embedding.}

\subsection{Publicly Reachable NAT64 Gateways}

\begin{table}[tb]
	\centering
	\begin{tabular}{llcccccc}
		\toprule
		& \textbf{Domain} & \textbf{github} & \textbf{azure} & \textbf{cloudflare} & \textbf{ntp} & \textbf{webex} \\
		\midrule
		\textbf{IPv6} & \textbf{IPs Measured} & 65 & 64 & 60 & 61 & 63 \\
		& \textbf{Correct Certificates} & 9 & 10 & 9 & 9 & 7 \\
		& \textbf{HTTP responses} & 14 & 17 & 4 & 4 & 13 \\
		\textbf{IPv4} & \textbf{IPs Measured} & 179 & 187 & 166 & 185 & 174 \\
		& \textbf{Correct Certificates} & 4 & 3 & 1 & 3 & 3 \\
		& \textbf{HTTP responses} & 7 & 6 & 3 & 3 & 6 \\
		\bottomrule
	\end{tabular}
	\caption{Results of our HTTP/S measurements for each domain. We highlight that a low number of NAT64 gateways are open on the Internet.}
	\label{table:zgrab_results}
\end{table}

We present the results of our HTTP and HTTPS measurements towards public NAT64 gateways in Table~\ref{table:zgrab_results}. 
From this analysis, we determine that there are a very low number of open NAT64 gateways on the Internet, though there are slightly more using IPv6 resolvers compared to IPv4. 
In IPv6, 6.6--26.6\% of HTTP responses and 11.1--15.6\% of HTTPS responses are ``correct.''
In IPv4, 1.6--3.9\% of HTTP responses and 0.6--2.23\% of HTTPS responses are ``correct.'' 

In our certificate analysis, we find that many of the ``incorrect'' certificates are from domain name providers, CDN providers, or default certificates for parked domains. 
We additionally identify other certificates from websites that have nothing to do with hosting or Internet infrastructure, such as a German medical device company. 
Finally, we find that one gateway consistently reflects back the certificate that we have on our measurement server. 
In our analysis of HTTP responses, we similarly find that ``incorrect'' responses largely consist of redirects to other hosting provider websites, parked domain websites, or other websites that have nothing to do with our measured domains (such as \texttt{netflix.com}).

For RIPE Atlas probes that answer with public IPv6 addresses, we once again measure the IPs for HTTP/S to see if the gateway is publicly reachable; see \Cref{table:ripe_atlas_zgrab_results}.
Compared to our results presented in \Cref{table:zgrab_results}, we are able to access more content through each IP returned by RIPE Atlas probes. 
We attribute this, the higher rates of correctly embedded addresses, and the stability across measurements to the nature of the RIPE Atlas measurement platform. 
As these probes are deployed explicitly for measurement, we hypothesize that there is more care around their connectivity. 
Therefore, if a probe uses NAT64, there is a higher likelihood that is configured correctly and consistently. 

\begin{table}[tb]
	\centering
	\begin{tabular}{@{}lllllll@{}}
		\toprule
		\textbf{Domain}           & \textbf{github} & \textbf{azure} & \textbf{cloudflare} & \textbf{ntp} & \textbf{webex}  \\
		\midrule
		\textbf{IPs Measured}     & 10              & 15             & 14                 & 13          & 10            \\
		\textbf{Correct Certificates} & 6               & 8              & 8                  & 7           & 6             \\
		\textbf{HTTP responses}       & 6               & 8              & 8                  & 7           & 6             \\
		\bottomrule
	\end{tabular}
	\caption{The results of our HTTP/S measurements for each domain with the answers from RIPE Atlas probes. We highlight that a low number of NAT64 gateways are open on the Internet.}
	\label{table:ripe_atlas_zgrab_results}
\end{table}

\takeaway{We find that NAT64 gateways located on public IPs are generally not publicly accessible.
The security implications of a gateway being open on the Internet can be severe, depending on whether it is intentional and where it is. 
We consider it a positive result that we find few open gateways.}
\section{Concluding Discussion}
\label{sec:conclusion}

While IPv6 deployment has certainly increased, IPv4 is not likely to disappear any time soon. 
Therefore, to maintain Internet connectivity, transition mechanisms will continue to be deployed. 

In this work, we presented an overview of NAT64 deployment. 
We measured public resolvers and included analysis of clients through our use of RIPE Atlas probes. 
We additionally used the RIPE Atlas probes that deploy NAT64 to measure towards our own web server.
Our measurements from three unique perspectives allowed us to understand NAT64 deployment from the resolver, client, and web server perspectives.
Specifically, we found that NAT64 resolvers are not deployed publicly at large, but there is notable concentration in China, mobile providers, and other network service providers.
Through our measurements of public resolvers and RIPE Atlas probes, we identified several open NAT64 gateways.
We emphasized the security implications if this was not purposeful, as they could be used for malicious intent or to access restricted content.
Finally, the relative AS of the NAT64 gateway, DNS64 resolver, and client using the transition protocol varies.

\parx{Future Work:}
We recommend that future work uses a variety of domains to measure the upper bound of NAT64 deployment because of the disparity in resolver responses.  
We note that an IPv6-to-IPv4 transition mechanism has no need to have an IPv4 address for the purposes of DNS64. 
We therefore hypothesize that the IPv4 resolvers that we find are dual-stack. 
This additionally implies that there are a substantial number of IPv6 resolvers that have not been found by the IPv6 hitlist~\cite{clusters_in_the_expanse}. 
With more robust IPv6 active address discovery techniques, it is possible that more NAT64 deployments could be discovered. 
Finally, we emphasize the need to understand how these mechanisms are deployed across client networks through the expansion of measurement platforms such as RIPE Atlas. 
However, we additionally highlight that these transition protocols likely exist largely in private networks that researchers have little vantage into, and therefore our measurements can only be taken as a lower bound.

\parx{Acknowledgments:}
This work was supported in part by National Science Foundation
(NSF) Graduate Research Fellowship (GRFP) under Grant No. DGE-2039655,
and NSF CNS award 2319315.

\bibliographystyle{splncs04}
\bibliography{refs}

\begin{thebibliography}{10}
\providecommand{\url}[1]{\texttt{#1}}
\providecommand{\urlprefix}{URL }
\providecommand{\doi}[1]{https://doi.org/#1}

\bibitem{transition_performance_2021}
Al-hamadani, A., Lencse, G.: {A survey on the performance analysis of IPv6
  transition technologies}. Acta Technica Jaurinensis  \textbf{14} (02 2021).
  \doi{10.14513/actatechjaur.00577}

\bibitem{464xlat_whitepaper}
{Alcatel Lucent}: {464XLAT in mobile networks IPv6 migration strategies for
  mobile networks} (2017),
  \url{https://www.apnic.net/wp-content/uploads/2017/01/IPv6_Migration_Strategies_for_Mobile_Networks_Whitepaper.pdf}

\bibitem{newer_rfc8215}
Anderson, T.: {Local-Use IPv4/IPv6 Translation Prefix}. RFC 8215 (Aug 2017).
  \doi{10.17487/RFC8215}, \url{https://www.rfc-editor.org/info/rfc8215}

\bibitem{apple_ipv6_req}
{Apple Developer Support}: {IPv6-Only Networks},
  \url{https://developer.apple.com/support/ipv6/}

\bibitem{ripe_atlas_docs}
Atlas, R.: Technical details,
  \url{https://atlas.ripe.net/docs/faq/technical-details.html}

\bibitem{464xlat_tmobile}
Byrne, C.: {464XLAT: Breaking Free of IPv4}. Presentation (2014),
  \url{https://conference.apnic.net/data/37/464xlat-apricot-2014\_1393236641.pdf}

\bibitem{6to4_rfc3056}
Carpenter, B.E., Moore, K.: {Connection of IPv6 Domains via IPv4 Clouds}. RFC
  3056 (Feb 2001). \doi{10.17487/RFC3056},
  \url{https://www.rfc-editor.org/info/rfc3056}

\bibitem{citizenlab_testlist}
{Citizen Lab and Others}: {URL testing lists intended for discovering website
  censorship} (2014), \url{https://github.com/citizenlab/test-lists}

\bibitem{dittrich2012menlo}
Dittrich, D., Kenneally, E.: {The Menlo Report: Ethical Principles Guiding
  Information and Communication Technology Research}. SSRN Electronic Journal
  (08 2012)

\bibitem{nat64.net}
Dupont, K.: {Public NAT64 service} (2019), \url{https://nat64.net/}

\bibitem{censys}
Durumeric, Z., Adrian, D., Mirian, A., Bailey, M., Halderman, J.A.: A search
  engine backed by internet-wide scanning. In: Proceedings of the 22nd ACM
  SIGSAC Conference on Computer and Communications Security. p. 542–553. CCS
  '15, Association for Computing Machinery, New York, NY, USA (2015).
  \doi{10.1145/2810103.2813703}, \url{https://doi.org/10.1145/2810103.2813703}

\bibitem{zmap}
Durumeric, Z., Wustrow, E., Halderman, J.A.: {ZMap: Fast Internet-wide Scanning
  and Its Security Applications}. In: 22nd USENIX Security Symposium (USENIX
  Security 13). pp. 605--620 (2013)

\bibitem{fortinet_dns}
{Fortinet}: D{NS over QUIC and DNS over HTTP3 for transparent and local-in DNS
  modes},
  \url{https://docs.fortinet.com/document/fortigate/7.4.1/administration-guide/8405/dns-over-quic-and-dns-over-https3-for-transparent-and-local-in-dns-modes-new}

\bibitem{clusters_in_the_expanse}
Gasser, O., Scheitle, Q., Foremski, P., Lone, Q., Korczy\'{n}ski, M., Strowes,
  S.D., Hendriks, L., Carle, G.: {Clusters in the Expanse: Understanding and
  Unbiasing IPv6 Hitlists}. In: Proceedings of the Internet Measurement
  Conference 2018. p. 364–378. IMC '18, Association for Computing Machinery,
  New York, NY, USA (2018). \doi{10.1145/3278532.3278564},
  \url{https://doi.org/10.1145/3278532.3278564}

\bibitem{transition_mechs_2005_rfc4213}
Gilligan, R.E., Nordmark, E.: {Basic Transition Mechanisms for IPv6 Hosts and
  Routers}. RFC 4213 (Oct 2005). \doi{10.17487/RFC4213},
  \url{https://www.rfc-editor.org/info/rfc4213}

\bibitem{google_public_dns64}
Google: {Google Public DNS64},
  \url{https://developers.google.com/speed/public-dns/docs/dns64}

\bibitem{google_ipv6_stat}
Google: {IPv6 Statistics} (2023),
  \url{https://www.google.com/intl/en/ipv6/statistics.html}

\bibitem{ietf_mailing_list_v6ops}
Howard, L.: {[v6ops] Transition mechanisms in use} (2018),
  \url{https://mailarchive.ietf.org/arch/msg/v6ops/\_8SKyRon\_tbZb4l1F9Ysly5ZGSM/}

\bibitem{iana_special_use}
IANA: I{ANA IPv6 Special-Purpose Address Registry} (2023),
  \url{https://www.iana.org/assignments/iana-ipv6-special-registry/iana-ipv6-special-registry.xhtml}

\bibitem{ieee_oui}
IEEE: {IEEE OUI} (2023), \url{https://standards-oui.ieee.org/oui/oui.txt}

\bibitem{zdns}
Izhikevich, L., Akiwate, G., Berger, B., Drakontaidis, S., Ascheman, A.,
  Pearce, P., Adrian, D., Durumeric, Z.: {ZDNS: A Fast DNS Toolkit for Internet
  Measurement}. In: Proceedings of the 22nd ACM Internet Measurement
  Conference. p. 33–43. IMC '22, Association for Computing Machinery, New
  York, NY, USA (2022). \doi{10.1145/3517745.3561434},
  \url{https://doi.org/10.1145/3517745.3561434}

\bibitem{plight_at_the_end_kristoff_21}
Kristoff, J., Ghasemisharif, M., Kanich, C., Polakis, J.: {Plight at the End of
  the Tunnel: Legacy IPv6 Transition Mechanisms in the Wild}. In: Passive and
  Active Measurement: 22nd International Conference, PAM 2021, Virtual Event,
  March 29 – April 1, 2021, Proceedings. p. 390–405. Springer-Verlag,
  Berlin, Heidelberg (2021). \doi{10.1007/978-3-030-72582-2\_23},
  \url{https://doi.org/10.1007/978-3-030-72582-2\_23}

\bibitem{tranco_LePochat2019}
{Le Pochat}, V., {Van Goethem}, T., Tajalizadehkhoob, S., Korczy\'{n}ski, M.,
  Joosen, W.: {Tranco: A Research-Oriented Top Sites Ranking Hardened Against
  Manipulation}. In: Proceedings of the 26th Annual Network and Distributed
  System Security Symposium. NDSS 2019 (Feb 2019).
  \doi{10.14722/ndss.2019.23386}

\bibitem{transition_classification_2019}
Lencse, G., Kadobayashi, Y.: {Comprehensive Survey of IPv6 Transition
  Technologies: A Subjective Classification for Security Analysis}. IEICE
  Transactions on Communications  \textbf{E102.B} (04 2019).
  \doi{10.1587/transcom.2018EBR0002}

\bibitem{rfc6144}
Li, X., Baker, F., Yin, K., Bao, C.: {Framework for IPv4/IPv6 Translation}. RFC
  6144 (Apr 2011). \doi{10.17487/RFC6144},
  \url{https://www.rfc-editor.org/info/rfc6144}

\bibitem{rfc6146_nat64}
Matthews, P., van Beijnum, I., Bagnulo, M.: {Stateful NAT64: Network Address
  and Protocol Translation from IPv6 Clients to IPv4 Servers}. RFC 6146 (Apr
  2011). \doi{10.17487/RFC6146}, \url{https://www.rfc-editor.org/info/rfc6146}

\bibitem{464xlat_rfc6877}
Mawatari, M., Kawashima, M., Byrne, C.: {464XLAT: Combination of Stateful and
  Stateless Translation}. RFC 6877 (Apr 2013). \doi{10.17487/RFC6877},
  \url{https://www.rfc-editor.org/info/rfc6877}

\bibitem{nat64_tayga_testing}
Opredelennov, E.: {NAT64 setup using tayga} (2016),
  \url{https://packetpushers.net/nat64-setup-using-tayga/}

\bibitem{partridge2016ethical}
Partridge, C., Allman, M.: {Ethical Considerations in Network Measurement
  Papers}. Commun. ACM  \textbf{59}(10),  58–64 (sep 2016)

\bibitem{ripe_atlas_platform}
{RIPE NCC}: {RIPE Atlas} (2023), \url{https://atlas.ripe.net/}

\bibitem{client_side_dns_schomp2013}
Schomp, K., Callahan, T., Rabinovich, M., Allman, M.: {On Measuring the
  Client-Side DNS Infrastructure}. In: Proceedings of the 2013 Conference on
  Internet Measurement Conference. p. 77–90. IMC '13, Association for
  Computing Machinery, New York, NY, USA (2013). \doi{10.1145/2504730.2504734},
  \url{https://doi.org/10.1145/2504730.2504734}

\bibitem{steger2023targetacquired}
Steger, L., Kuang, L., Zirngibl, J., Carle, G., Gasser, O.: {Target Acquired?
  Evaluating Target Generation Algorithms for IPv6}. In: Proceedings of the
  Network Traffic Measurement and Analysis Conference (TMA) (Jun 2023)

\bibitem{6to4_anycast_deprecated_rfc7526}
Trøan, O., Carpenter, B.E.: {Deprecating the Anycast Prefix for 6to4 Relay
  Routers}. RFC 7526 (May 2015). \doi{10.17487/RFC7526},
  \url{https://www.rfc-editor.org/info/rfc7526}

\bibitem{employees_aaaa_stats}
Wing, D.: {AAAA and IPv6 Connectivity statistics} (2023),
  \url{https://www.employees.org/~dwing/aaaa-stats/}

\bibitem{china_rfc6219}
Wu, J., Zhang, H., Li, X., Chen, M., Bao, C.: {The China Education and Research
  Network (CERNET) IVI Translation Design and Deployment for the IPv4/IPv6
  Coexistence and Transition}. RFC 6219 (May 2011). \doi{10.17487/RFC6219},
  \url{https://www.rfc-editor.org/info/rfc6219}

\bibitem{zirngibl2022rusty}
Zirngibl, J., Steger, L., Sattler, P., Gasser, O., Carle, G.: {Rusty Clusters?
  Dusting an IPv6 Research Foundation}. In: ACM Internet Measurement Conference
  2022 (Oct 2022). \doi{10.1145/3517745.3561440}

\bibitem{zgrab2}
ZMap: {ZGrab2} (2023), \url{https://github.com/zmap/zgrab2}

\bibitem{nat64_checker}
Zorz, J.:  (2017),
  \url{https://labs.ripe.net/author/janzorz/introducing-nat64-checker/}

\end{thebibliography}

\appendix
\section{Appendix}

\subsection{Details on Resolver Responses}
\label{app:resolver_responses}
\begin{figure}
\includegraphics[width=\columnwidth]{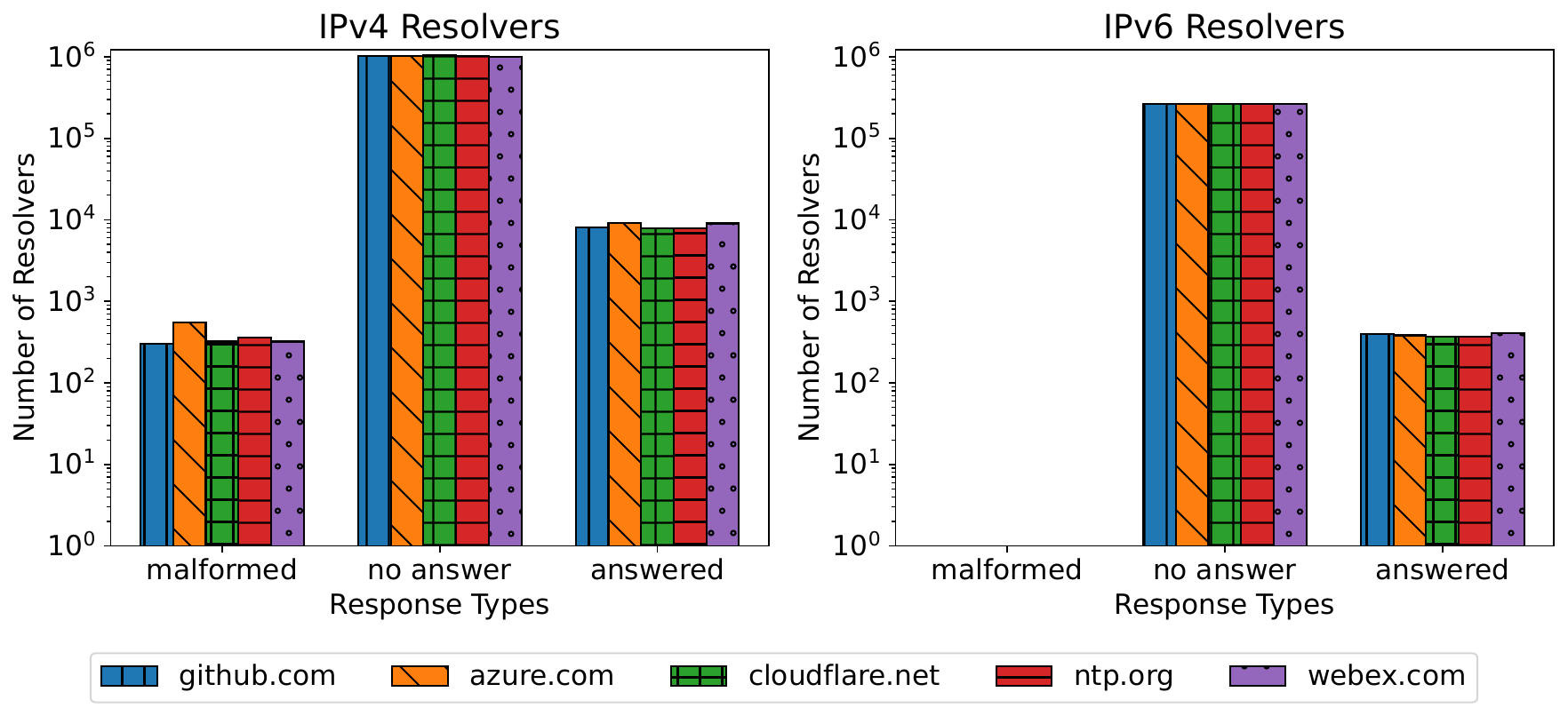}
\caption{An overview of the types of DNS responses to our measurements. For DNS responses with no answer, the \texttt{rcode} is always NOERROR.
}
\label{fig:overview_resolvers}
\end{figure}

Figure~\ref{fig:overview_resolvers} gives an overview of the types of responses the resolvers sent to our queries. 
We see that the queries get overall similar numbers of response types, independent of the queried domain.

\begin{figure}
\includegraphics[width=\columnwidth]{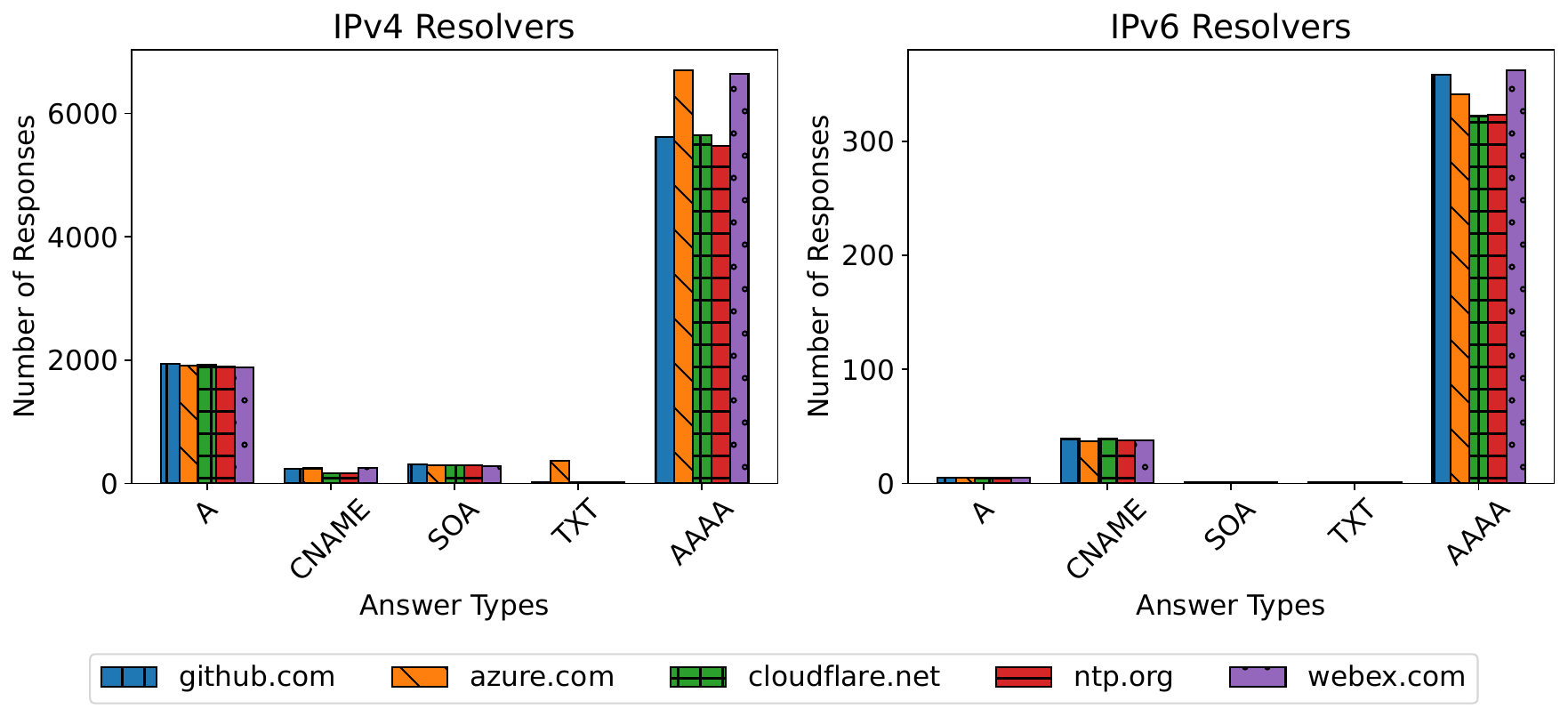}
\caption{An overview of the answer types in the resolver responses across domains.
We note that from IPv4 resolvers we also receive less than 10 NS records, less than 15 HINFO records, 1 MX record, less than 4 OPT responses, and less than 3 RRSIG responses. 
From IPv6 resolvers, we also receive one HINFO response. 
For brevity, we omit these values from the Figure. 
}
\label{fig:overview_resolver_answers}
\end{figure}

Figure~\ref{fig:overview_resolver_answers} shows an overview of the resource record types of answers the resolvers sent for each domain. 

\subsection{Resolver Answers}
\label{app:resolver_answers}
\begin{figure}
	\includegraphics[width=\columnwidth]{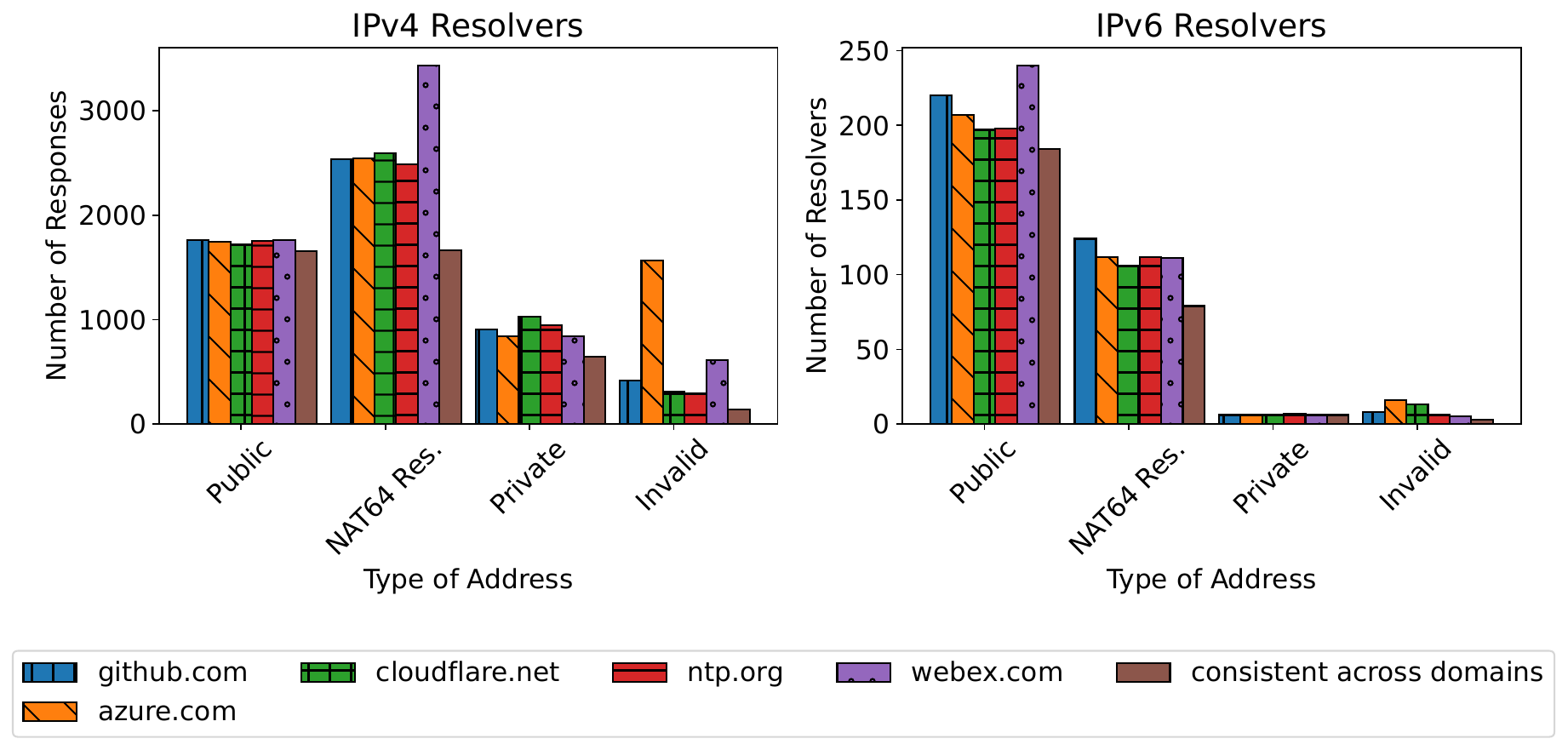}
	\caption{An overview of the types of IPv6 addresses on the AAAA records returned. 
		We count one response per resolver.
		``NAT64 Res.'' exclusively refers to the NAT64 special use prefix. 
		``Private'' is other private IPv6 address space.
		``Invalid'' is an IPv6 address that is not in a valid format (\ie starts with `::').
		The bar labeled ``consistent across domains'' refers to the resolvers that occur in every domain measurement in the respective address type. 
	}
	\label{fig:resolver_aaaa}
\end{figure}

Although Figure~\ref{fig:overview_resolvers} shows a relatively uniform distribution in the types of responses across resolvers, we see that certain domains elicit different types of responses across resolvers in Figure~\ref{fig:overview_resolver_answers}.
We focus on AAAA answers and the differences in their resolver responses. 
Figure~\ref{fig:resolver_aaaa} shows the distribution of the addresses in AAAA answers received from resolvers. 

In the left subfigure of Figure~\ref{fig:resolver_aaaa} across all domain measurements (the brown bar), 1,663 always respond with a public IPv6 address, 1,664 always respond with an address in a special-use prefix as defined by IANA~\cite{iana_special_use} (\eg documentation prefix or private address prefix), 141 always respond with an invalid IPv6 address in the answer field (\ie the \texttt{ANCOUNT} field is 1 but the answer field is populated with \texttt{::}, or an address that starts with \texttt{::}), and 635 always respond with an answer in a different private prefix. 

In Figure~\ref{fig:resolver_aaaa} IPv6 resolvers, 185 always respond with a public IPv6 address, 79 with an address in the special-use prefix, 3 with an empty answer, and 5 with an address in another private prefix.

We highlight the high number of IPv4 resolvers that answer with the special-use prefix for \texttt{webex.com}. 
There are 895 resolvers that always answer for \texttt{webex.com} that answer inconsistently for the other domains, but when they do, they use the NAT64 prefix. 
Additionally, there are 118 resolvers that answer for other domains with an `ancount` of 0 (\ie with no answer). 
Finally, 684 resolvers do not respond at all for other domains. 
Of the resolvers that respond for \texttt{webex.com}, 92.8\% are concentrated in AS 4538, China Education Network. 
An additional 19.7\% are in AS 134540 (Tata Teleservices). 
We hypothesize that these networks use WebEx as a service and, therefore, this domain over others.
There are also higher response rates for \texttt{azure.com} where the answer fields are invalid in IPv4 resolvers. 
1,168 of these resolvers only respond to queries for \texttt{azure.com},
43.0\% of which are in AS 9808, China Mobile Network, and 28.9\% of which are in AS 4837, China Unicom Backbone. We cannot attribute these abnormalities to any property in particular but note that prior work also finds irregularities in Chinese resolvers on requests for AAAA records~\cite{steger2023targetacquired}.

Additionally, more IPv6 resolvers answer with public IPs for \texttt{webex.com} than other domains. 
All of these are in AS20940, Akamai, and all resolvers answer with IP \texttt{2001:4801:7829:105:be76:4eff:fe10:1fc0}, an EUI-64 address that encodes a MAC address. 
The Organizational Unique Identifier (OUI) in this address belongs to Rackspace, US, Inc., a cloud computing company~\cite{ieee_oui}.

In our filtering process described in Section~\ref{sec:dataset},
we find that 2,454 IPv4 resolvers always answer with the same value across different domain measurements.
Of these, 991 always respond with \texttt{2620:101:9000:53::55}, which is used as a redirect address in FortiGuard with DNS over HTTP3~\cite{fortinet_dns}.
449 also always respond with \texttt{2607:f740:100:108::1400} and 103 always respond with an invalid IPv6 address in the answer field.
172 IPv6 resolvers always answer with the same value in the AAAA answer field across different domain measurements. 
In 114 of these cases, the answer is \texttt{2607:f740:100:108::1400}, an IP address allocated to NetActuate, a network services and infrastructure provider.
We hypothesize that this is also a default answer for the resolver software from this company or a redirect. 
 \end{document}